\documentclass[a4paper]{jpconf}
\pdfoutput=1
\usepackage{graphicx}
\usepackage{caption}
\usepackage{subcaption}
\usepackage{array}
\usepackage{slashed}
\usepackage{amsmath}
\usepackage{amssymb}

\usepackage[square,sort&compress,numbers]{natbib}
\def\trm{\tr_-}
\def\trp{\tr_+}
\def\la{\langle}
\def\ra{\rangle}
\def\spA#1#2{\la#1#2\ra}

\begin{document}

\title{Recent progress for five-particle two-loop scattering amplitudes with an off-shell leg}

\author{Jakub Kry\'s}

\address{Dipartimento di Fisica and Arnold-Regge Center, Università di Torino, and INFN, Sezione di Torino, Via P. Giuria 1, I-10125 Torino, Italy}
\address{Institute for Particle Physics Phenomenology, Department of Physics, Durham University, Durham DH1 3LE, United Kingdom
}

\ead{jakubmarcin.krys@unito.it}

\begin{abstract}
We report on the advances in the calculation of the two-loop scattering amplitudes for five-particle processes with one off-shell leg. Focusing on the production of a Higgs boson in association with a bottom quark pair, we outline how the newly developed technology allows us to overcome the computational bottlenecks. In particular, we discuss the use of finite field arithmetic and elucidate a convenient way to evaluate numerically the special functions appearing in the amplitudes.
\end{abstract}

\section{Introduction}
The growing need for increasingly precise theoretical predictions of LHC observables has challenged the community to tackle processes of enormous algebraic and analytic complexity. The main obstacles are the construction of master integrals satisfying canonical differential equations (DEs) and the determination of their rational coefficients. High-multiplicity final states constitute a key ingredient of QCD at NNLO accuracy. In recent years, calculation of scattering amplitudes of certain $2\to3$ processes has been made possible by rapid progress in computational technology. In particular, the use of finite-field arithmetic has emerged as a valuable tool for overcoming the complexity of intermediate steps by reconstructing the analytic coefficients from their numerical evaluations only at the final stage~\cite{vonManteuffel:2014ixa,Peraro:2016wsq,Klappert:2019emp,Peraro:2019svx,Klappert:2020aqs,Klappert:2020nbg}. However, the determination of a canonical master integral basis remains a major challenge. For $2\to3$ massless scattering, results are known including the non-planar integrals~\cite{Papadopoulos:2015jft,Gehrmann:2018yef,Abreu:2018rcw,Chicherin:2018mue,Chicherin:2018old,Chicherin:2020oor}. Introduction of one massive leg further complicates the construction of such basis. Nonetheless, results for the planar sector are known~\cite{Papadopoulos:2015jft,Abreu:2020jxa,Canko:2020ylt,Syrrakos:2020kba} and enabled the recent computation of the $b\bar{b}H$ production amplitudes within the leading colour approximation. The full colour computation requires results for one-mass non-planar integrals, where only the hexa-box families are known~\cite{Papadopoulos:2019iam,abreu2021twoloop} so far\footnote{See also~\cite{Kardos:2022tpo}, released after our publication.}, with results for the non-planar double-pentagon families still missing. Overall, canonical DEs have already proven extremely useful in calculations of high-multiplicity processes~\cite{Henn:2013pwa}. Moreover, the method of evaluating master integrals by exploiting their DEs has been widely adopted. Indeed, the use of generalised series expansions to solve DEs has enabled stable and reliable numerical evaluation of amplitudes across the full phase space~\cite{Francesco:2019yqt,Hidding:2020ytt}.

In these proceedings, we summarise the recent computation of the two-loop leading-colour QCD helicity amplitudes for Higgs boson production in association with a bottom-quark pair~\cite{Badger:2021ega}. This process enables the direct measurement of the Yukawa coupling between bottom quarks and the Higgs boson, which can be enhanced according to certain supersymmetric scenarios~\cite{Balazs:1998nt,Dawson:2005vi}, hence offering a valuable test of the Standard Model parameter.

\section{Amplitudes for $b\bar{b}H$ production}

\begin{figure}
	\begin{center}
		\includegraphics[width=0.95\textwidth]{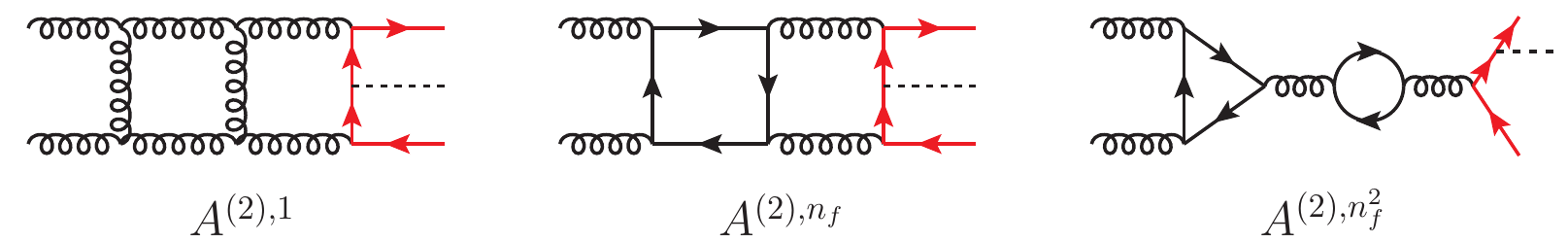}
	\end{center}
	\caption{Sample Feynman diagrams corresponding to the various closed fermion loop contributions at two loops as specified in Eq.~\eqref{eq:NfDecomposition2L}.
		Black-dashed, red, black-spiralled and black lines represent Higgs bosons, bottom quarks, gluons and light quarks (bottom quarks included), respectively.}
	\label{fig:amp2L}
\end{figure}

We compute the two-loop QCD scattering amplitudes for the following processes:
\begin{align} \label{eq:subprocesses}
	& 0 \rightarrow \bar{b}(p_1) + b(p_2) + g(p_3) + g(p_4) + H(p_5)\,, \\
	& 0 \rightarrow \bar{b}(p_1) + b(p_2) + \bar{q}(p_3) + q(p_4) + H(p_5)\,, \\
	& 0 \rightarrow \bar{b}(p_1) + b(p_2) + \bar{b}(p_3) + b(p_4) + H(p_5)\,,
\end{align}
with $p_1^2=p_2^2=p_3^2=p_4^2=0\,$ and $p_5^2=m_H^2\,$. We work within the leading colour and massless $b$-quark approximations. The five-particle kinematics is parametrised by the following Mandelstam variables:
\begin{equation}
	\vec{s}_5 = \{s_{12},s_{23},s_{34},s_{45},s_{15},p_5^2\} \,, \nonumber
\end{equation}
while the parity information is fully described by the pseudo-scalar invariant:
\begin{equation} \label{eq:kinematics}
	\text{tr}_5 = 4 i \epsilon_{\mu \nu \rho \sigma} p_1^{\mu} p_2^{\nu} p_3^{\rho} p_4^{\sigma} = [12] \spA{2}{3} [34] \spA{4}{1} - \spA{1}{2} [23] \spA{3}{4} [41] \,.
\end{equation}
After colour decomposition, the partial amplitudes are further decomposed according to the fermion content:
\begin{align}
	& A^{(1)} = N_c A^{(1),1} + n_f A^{(1),n_f}  \,,
	\label{eq:NfDecomposition1L} \\
	& A^{(2)} = N_c^2 A^{(2),1} + N_c n_f A^{(2),n_f} + n_f^2 A^{(2),n_f^2}  \,,
	\label{eq:NfDecomposition2L}
\end{align}
where $n_f$ is the number of light quarks circulating in the loop. Sample Feynman diagrams contributing to the process are shown in Fig.~\ref{fig:amp2L}.
The $L$-loop finite remainders of the partial amplitudes can then be obtained by subtracting the pole operators which capture the complete IR and UV singularity structure~\cite{Catani:1998bh}:
\begin{equation}
	F^{(L)} = \lim_{\epsilon \to 0} \left[ A^{(L)} - P^{(L)} A^{(0)} \right]\,.
	\label{eq:finiteremainder}
\end{equation}

\section{Reduction and reconstruction}
We use Feynman diagrams as the starting point of our calculation. Following colour decomposition, each helicity amplitude is expressed as a sum of integrals corresponding to various topologies. The numerators of these integrals contain monomials involving loop momenta, with coefficients that depend on external kinematics only. Each topology is then mapped onto one of 15 independent maximal topologies and the loop monomials in the corresponding numerators are expressed in terms of the propagators of the target topology. In order to avoid the huge analytic complexity, at this stage we begin the numerical sampling procedure. To facilitate the use of finite-field arithmetic, we express the kinematic variables of Eq.~\eqref{eq:kinematics} in terms of momentum-twistor variables~\cite{Hodges:2009hk,Badger:2016uuq}. The following parametrisation:
\begin{equation}
	\begin{aligned}
		x_1 &= s_{12} && \qquad x_4 = \frac{s_{23}}{s_{12}}, \\
		x_2 &= -\frac{\trp(1234)}{s_{12}s_{34}} && \qquad x_5 = -\frac{\trm(1(2+3)(1+5)523)}{s_{23}\trm(1523)}, \\
		x_3 &= \frac{\trp(134152)}{s_{13}\trp{(1452)}} && \qquad x_6 = \frac{s_{45}}{s_{12}}, \\
	\end{aligned}
	\label{eq:mtvardefs}
\end{equation}
where $\tr_{\pm}(ij \cdots kl) = \frac{1}{2} \tr((1\pm\gamma_5)\slashed{p}_i\slashed{p}_j \cdots\slashed{p}_k\slashed{p}_l)$, rationalises $\text{tr}_5$ and yields $x_1$ as the only dimensionful variable. This means that we can set it to be equal to 1, thereby simplifying the reconstruction process. The dependence of the amplitude on $x_1$ can be restored through dimensional analysis.

Scalar integrals obtained at the end of this mapping procedure can be further reduced onto a basis of independent master integrals through integration-by-parts reduction. We generate the system of IBP relations in \texttt{LiteRed}~\cite{Lee:2012cn} and then solve it numerically over finite-fields using the Laporta algorithm~\cite{Laporta:2001dd,Peraro:2019svx}. We further expand the master integrals into the special function basis constructed in Ref.~\cite{Badger:2021nhg}. By Laurent expanding the amplitude and subtracting the pole operators, we obtain the finite remainders written as a linear combination of monomials $m_i$ of special functions with rational coefficients of momentum-twistor variables $x$:
\begin{equation}
	F^{(L)} = \sum_{i} r_i(x) m_i(f)\,.
	\label{eq:finrem}
\end{equation}
To improve the reconstruction procedure of the coefficients, we employ several additional strategies. First, we find linear relations between the coefficients and reconstruct only the independent ones, which are purposefully chosen as the ones with the lowest possible polynomial degree. We then make use of an ansatz of factors which may appear in the denominators of the coefficients. The guess for the ansatz is inspired by the structure of the differential equations satisfied by the master integrals and by considering the singularities of the special functions. This allows us to fully determine the denominator factors, as well as some in the numerator. Finally, to further lower the degrees of the remaining expression that cannot be matched to the ansatz, we employ an in-house univariate partial fractioning procedure. We find that these tools lead to a drastic drop in the degrees of the rational coefficients and make	 their reconstruction feasible, as can be seen in Table~\ref{tab:degrees2q2bH}.
\begin{table}[t!]
	\caption{\label{tab:degrees2q2bH} Maximum numerator/denominator polynomial degrees of the finite remainder coefficients $r_i(x)$ in Eq.~\eqref{eq:finrem}	at each stage of our reconstruction steps, together with the number of sample points needed for the analytic reconstruction in the $\bar{b}b\bar{q}qH$ subprocess, for the various closed fermion loop contributions.}
	\begin{center}
	\begin{tabular}{m{1.0cm}m{2.3cm}m{1.0cm}m{2.0cm}m{2.3cm}m{1.8cm}}
		\br
		$\bar{b}b\bar{q}qH$     & helicity \newline configurations & $r_i(x)$ & independent $r_i(x)$ & partial \newline fraction in $x_5$ & number of \newline points \\
		\mr
		$F^{(2),1}$ & $+++-$ & 82/81 & 69/70 & 24/16 & 10326 \\
		$F^{(2),n_f}$ & $+++-$ & 28/30 & 25/24 & 8/6 & 379 \\
		$F^{(2),n_f^2}$ & $+++-$ & 6/11  & 6/11  & 3/0 & 22 \\
		\br
	\end{tabular}
	\end{center}
\end{table}
\section{A custom basis of special functions for the finite remainder}
The function space of two-loop one-mass amplitudes is larger than the one of the corresponding finite remainders for a specific process. For instance, as reported previously by several authors, the pole subtraction results in some of the letters of the differential equation alphabet dropping out of the expressions~\cite{Badger:2018enw,Abreu:2018zmy,Abreu:2018aqd,Chicherin:2018yne,Chicherin:2019xeg,Abreu:2019rpt,Abreu:2019odu,Badger:2019djh,Caron-Huot:2020vlo,Abreu:2020cwb,Chawdhry:2020for}. 
Therefore, we find it convenient to construct a modified version of the basis presented in Ref.~\cite{Badger:2021nhg} that is tailored to the finite remainders of the $b\bar{b}H$ production. We do this by selecting only those linearly independent combinations of special functions that appear in the finite remainder. In comparison with the original basis, we observe a drastic drop (from 113 to 23) in the number of required weight-4 functions. This allows us to significantly reduce the evaluation time of the finite remainders when using the generalised series expansion approach.

The basis elements are expressed in terms of Chen's iterated integrals~\cite{Chen:1977oja,Brown:2013qva}. These objects feature several properties that are particularly useful to our study. Crucially, they automatically implement functional relations that are otherwise hidden in other representations of special functions, such as Goncharov polylogarithms. This means that the aforementioned cancellations take place analytically rather than only at the stage of the numerical evaluation of the results. Moreover, we use them to construct the DEs satisfied by the new, process-specific basis, which we then solve in \texttt{DiffExp}~\cite{Hidding:2020ytt} to evaluate the finite remainders.
\section{Results}
\begin{figure}[b!]
	\centering
	\begin{subfigure}{.5\textwidth}
		\centering
		\includegraphics[width=.9\textwidth]{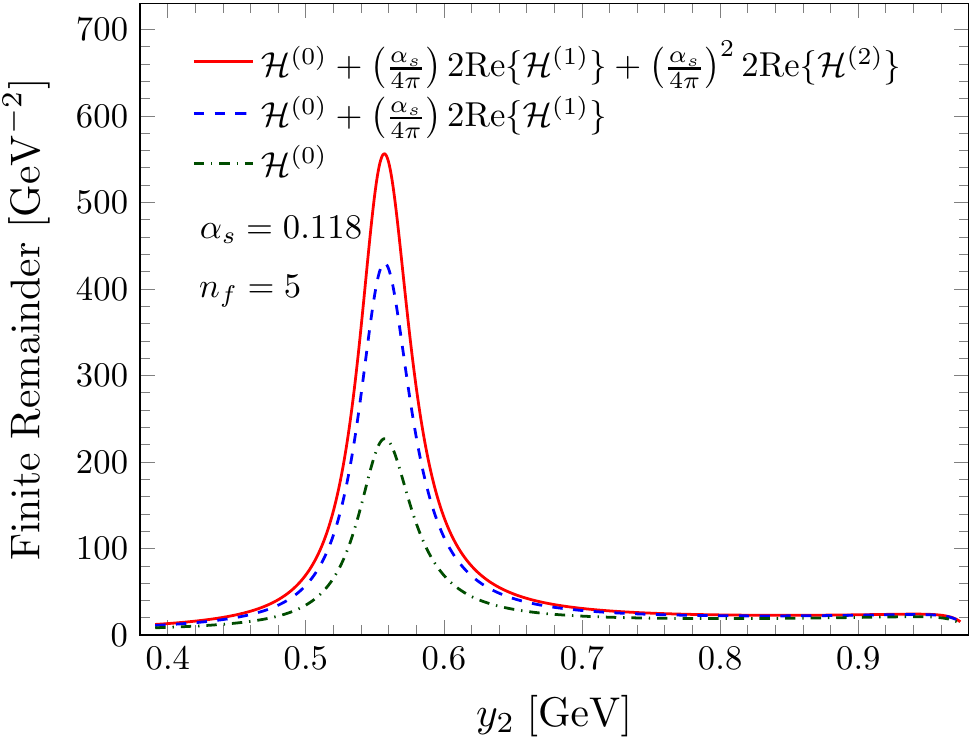}
		\label{fig:gg}
		\caption{$\mathbf{gg}$}
	\end{subfigure}%
	\begin{subfigure}{.5\textwidth}
		\centering
		\includegraphics[width=.9\textwidth]{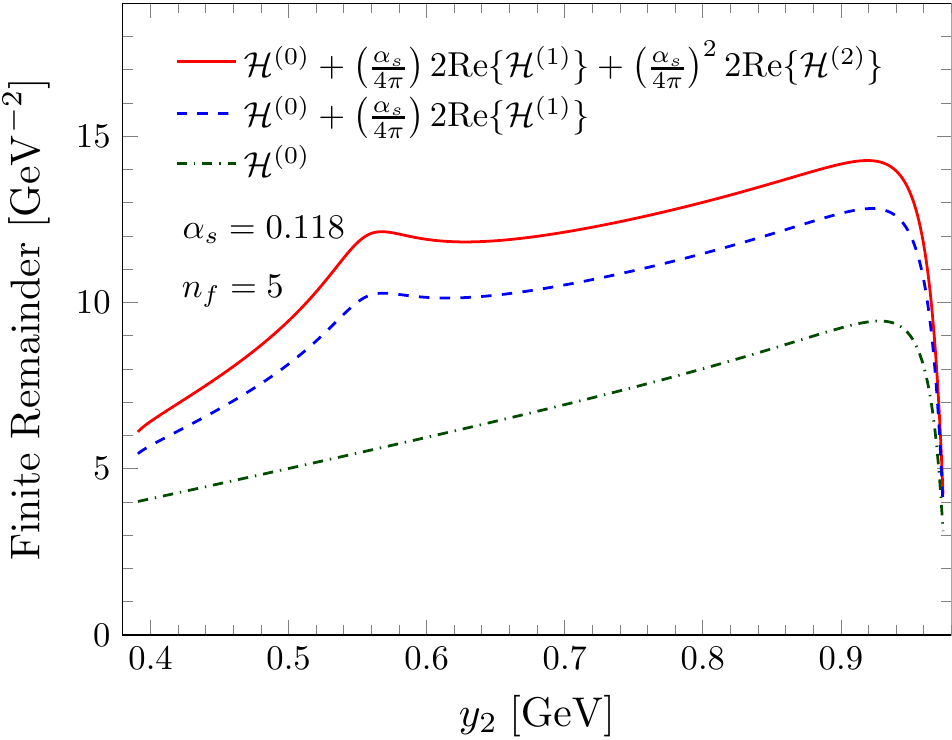}
		\label{fig:qxq}
		\caption{$\mathbf{\bar{q}q}$}
	\end{subfigure}%
	
	\caption[short]{Reduced squared finite remainders $\mathcal{H}^{(L)}$ at tree level, one and two loops evaluated on an arbitrarily chosen one-dimensional phase space slice for the $gg$ and $\bar{q}q$ channels.}
	\label{fig:plots}
\end{figure}
We verify the validity of our results in several ways, i.e. by checking the dependence of the finite remainders on the renormalisation scale, confirming a non-trivial relation between expressions for two helicity configurations and studying convergence of the finite remainders near their spurious poles. To demonstrate the suitability of our method for phenomenological applications, we present results for the finite remainders interfered with tree-level amplitudes, evaluated on an arbitrary univariate phase space slice. Example plots for the $gg$ and $\bar{q}q$ channels are shown in Fig.~\ref{fig:plots}. We also compute the tree-level expressions needed to obtain the squared finite remainders using BCFW recursion relations~\cite{Britto:2004ap,Britto:2005fq}. Their analytic form helps elucidate the features of the plots. For instance, the peaks in the $gg$ channel are due to the $s_{23} = 0$ poles in the tree-level amplitudes, while the loop-induced peaks in the $\bar{q}q$ channel can be traced back to logarithms of $s_{24}$ present in the finite remainder special functions.
Overall, the techniques presented here show great promise for applications to other important scattering processes. In fact, they have already been used to compute the two-loop QCD helicity amplitudes for $W^\pm\gamma+j$ production within the leading colour approximation~\cite{Badger:2022ncb}. We look forward to the experimental tests of NNLO QCD theoretical predictions that will become available in the near future.

\ack
JK would like to thank the organisers of the ACAT 2021 conference for the opportunity to present this work. This project received funding from the European Union's Horizon 2020 research and innovation programmes
\textit{New level of theoretical precision for LHC Run 2 and beyond} (grant agreement No 683211),
\textit{High precision multi-jet dynamics at the LHC} (grant agreement No 772009), and
\textit{Novel structures in scattering amplitudes} (grant agreement No 725110).

\bibliography{main}

\end{document}